\documentclass[%
 aip,
 amsmath,amssymb,
 reprint,%
]{revtex4-1}

\usepackage{graphicx}
\usepackage{dcolumn}
\usepackage{bm}

\usepackage[utf8]{inputenc}
\usepackage[T1]{fontenc}
\usepackage{mathptmx}
\usepackage{etoolbox}
\usepackage{epsfig}
\usepackage{amssymb}
\usepackage{color}
\usepackage{graphicx,epsfig}
\usepackage{xcolor}

\makeatletter
\def\@email#1#2{%
 \endgroup
 \patchcmd{\titleblock@produce}
  {\frontmatter@RRAPformat}
  {\frontmatter@RRAPformat{\produce@RRAP{*#1\href{mailto:#2}{#2}}}\frontmatter@RRAPformat}
  {}{}
}%
\makeatother
\begin{document}

\preprint{AIP/123-QED}

\title{Shaping the aggregates of discotic particles with directional pair interactions}
\author{Bruno Mart\'{\i}nez-Haya}
\author{Neftal\'{\i} Morillo}%
\author{Alejandro Cuetos}\email{acuemen@upo.es}
\affiliation{Center for Nanoscience and Sustainable Technologies (CNATS), and  Department of Physical, Chemical and Natural Systems, Universidad Pablo de Olavide, 41013 Seville, Spain.}

\date{\today}

\begin{abstract}
Aggregation processes in systems of planar macromolecules and colloids drive a broad range of phenomena in natural systems and soft materials. Depending on chemical architecture, intermolecular interactions in these systems may favor different relative pair orientations, such as stacking face--face or percolating edge--edge arrangements. In this work, we employ a versatile coarse-grained interaction model for disk-like particles to provide a general framework to rationalize the thermotropic formation of aggregates and predict the topology of the resulting suprastructures. Monte Carlo and Brownian Dynamics simulations show that, with an appropriate tuning of the interactions, discotics spontaneously nucleate into  clusters with globular, planar or stacked geometries, leading to materials with specific internal order and associated physicochemical properties. 
\end{abstract}

\maketitle

\section{\label{sec1}Introduction}

Discotic particles (platelets) have been a central topic in Soft Matter Science as paradigmatic benchmarks for steric interactions efficiently promoting liquid crystal positional and orientational ordering \cite{WOH16,KUM16,LAS07}. Hand-in-hand with the fundamental advances in the knowledge of their physicochemical behavior, discotic materials are nowadays incorporated into a diversity of technological applications, ranging from nanoelectronics, to photonics and sensor devices \cite{SER07,STO16}. 

Discotics stand out for their propensity to stack in columnar arrangements. As noted in a 2002 review article \cite{BUS02}, columnar assemblies prevail at very different length scales, ranging from ‘tens of angstroms (molecular), tens of nanometers (macromolecular or supramolecular), hundreds of nanometers (colloidal) or tens of microns (“manufactured” platelets)’ \cite{GOW18}. The directional overlap of molecular orbitals in columnar configurations leads to unique electronic functionalities that have been implemented in molecular wires, photovoltaic substrates, organic light-emitting diodes and other semiconductor devices \cite{BUS11,KUM06,OHT03,LAS07,SCH01}. 

More recent investigations have demonstrated that the appropriate tailoring of the pair interactions is capable of inducing a rich variety of partially ordered mesophases, including parallel, tilted and biaxial discotic columnar phases, but also less common discotic smectic, cubatic and uniaxial phases \cite{MAR10,MOR21}. 

The present study seeks further insights into the interrelation between the interactions and the collective organization of discotic particles. Complementing the above-mentioned investigations on liquid crystal phases, we focus here specifically on the low density regime, with the aim to characterize the onset of aggregation of the particles upon cooling, and the growth of anisotropic clusters. The aggregation of discotic molecular or colloidal structures has been identified as driving macromolecular collapse in organic and biological contexts ($e.g.$, in the flocculation of crude oil, or the formation of stacks of erythrocytes or amyloid plaques \cite{MUL08,NEH18,WIL21}). Our working hypothesis is that the directionality of the pair interactions largely determines the topology of the aggregated suprastructures, potentially yielding distinct electronic, optical and physico-chemical properties.  Face-face stacking interactions are most common in discotic colloids and drive fundamental aggregation processes in inorganic materials such as smectite clay nanoparticles \cite{SHE16}, complex organic polyaromatics \cite{MUL11}, or bioorganic b-sheet amyloid protein configurations \cite{GAZ02}. In a broader scope, the tayloring of colloidal particles with specific interacting groups is an active field of research. For instance, the use of biomolecular moieties (DNA or protein labeling) has been used for more than two decades to stabilize supramolecular arrangements in the nano and mesoscale \cite{MIR96}. The introduction of specific binding moieties may lead to edge-edge interactions the modulate or even dominate stacking interactions of discotics. Examples of edge-edge interacting groups, occasionally tuned by pH and ionic strength, can be found in polyaromatic sheets with aryl-dipole side interactions \cite{ALS01} and in the design of surfactant nanodiscs \cite{MAN11} or FEBEX bentonite colloids \cite{SEH20}. It is however not straightforward to draw precise predictions about aggregation processes from pair interactions alone, as the balance of steric interactions is subtle in discotics \cite{MOR21}. 

\begin{figure*}
	\centering
	\includegraphics[width=0.9\linewidth]{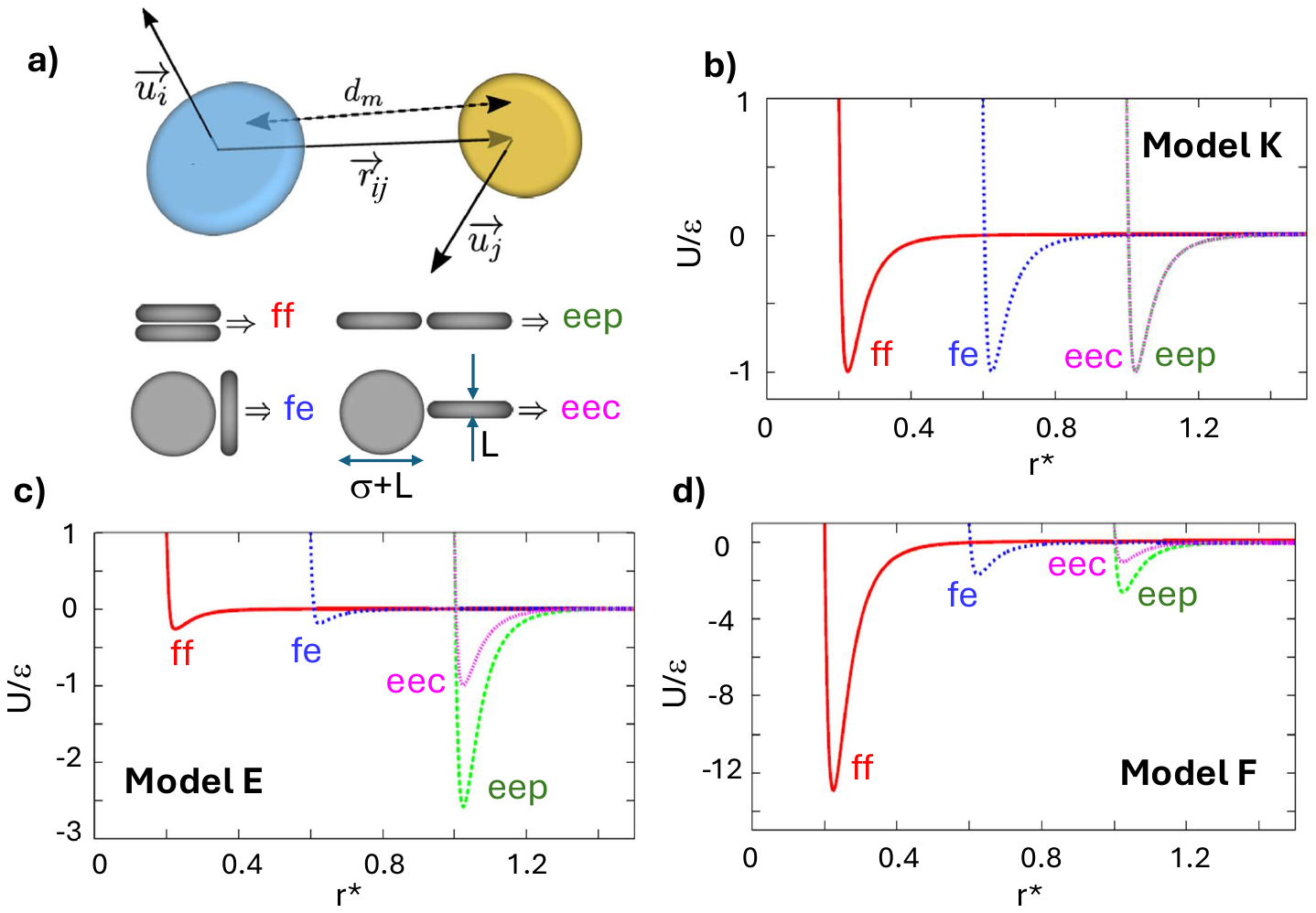}
	\vspace*{-1.6cm}
	\caption{\label{fig:ogbk} a) Illustration of the OGBK model for two interacting platelets \cite{MAR11,MOR21}. The interaction energy depends on the relative position of the particles (${\bf r}_{ij}$), director vectors that define the  orientation of the particles (${\bf \hat{u}}_i$ and ${\bf \hat{u}}_j$), and on the minimum distance between the particle cores ($d_m$) \cite{CUE08}. Four pair configurations are considered as reference to describe the model: stacking "face-face" (ff), T-shaped "face-edge" (fe), parallel "edge-edge" (eep), crossed "edge-edge" (eec). b--d): Potential energy curves for two platelets in each of the four reference configurations, interacting with the OGBK parametrization employed in this work: b) model K (uniform interactions), c) model E (eep favored) and d) model F (ff stacking preferred).}    
\end{figure*}

Simulation studies of discotic fluids have been less prolific than those dedicated to rod-like particle fluids \cite{ALL93,MED14,CAR05,WIL05,FRE85,VER92,ZAN99,CAP03,CHA08}. In recent years, the use of coarse-grained models has benefited from the development of efficient algorithms to compute interparticle distances and pair interactions \cite{WOJ84,CUE08,MAR09,MAR11}. Among many other relevant contributions, Carlos Vega was a pioneer in this field; his seminal work on fluids of discotic Kihara particles \cite{GAM08} has been a source of inspiration for the later developments in our group. 
Motivated by the general problem of modeling soft discotic materials, a versatile coarse-grained pair interaction potential was introduced in previous works, namely the Oblate-Gay-Berne-Kihara (OGBK) model \cite{MAR09,MAR10,MOR21}. The OGBK model features a facile modulation of the directionality of the pair interactions \cite{MOR21}. It is here employed to build fluids either with homogeneous interactions around the discotic core, or with interactions favoring either face-face or edge-edge pair configurations. This approach allows us to explore how directionality in particle interactions influences low-density aggregation phenomena and its role in the geometry and structure of the clusters formed. To this end, Monte Carlo and Brownian Dynamics simulations are employed to characterize thermotropic aggregation processes in each fluid model. The evolution of the internal structure of the fluid upon cooling is monitored with a focus on cluster size and shape. It is shown that this approach provides valuable insights into the possibility of a controlled tailoring of discotic aggregation phenomena.

\section{Methods and tools}

\subsection{\normalsize Interaction models}

The present study follows recent developments of the Oblate-Gay-Berne-Kihara model \cite{MAR09,MAR10,MOR21}. The OGBK interaction model is based on a combination of the Kihara core interaction potential and the directional interaction strengths of the Gay-Berne potential ($U_K$ and $\epsilon_{OGB}$ below, respectively). The geometrical parameters of the model are depicted graphically in Fig.\,\ref{fig:ogbk}. The OGBK interaction, $U_{OGBK}$, is mathematically described by the following set of equations:

\begin{equation}\label{OGBK} 
U_{OGBK}({\bf r_{ij}},{\bf \hat{u}_i},{\bf\hat{u}_j})
= \epsilon_{OGB} ({\bf \hat{r}_{ij}},{\bf \hat{u}_i},{\bf
	\hat{u}_j}) \, U_{K}(d_m)
\end{equation}
\begin{equation}\label{eq2}
U_{K} (d_m)=  4 \epsilon_0 \left[  \left( L/d_m \right)^{12} -
\left( L/d_m \right)^6 	\right] 
\end{equation}
\begin{equation}\label{OGB}
\epsilon_{OGB} ({\bf \hat{r}_{ij}},{\bf \hat{u}_i},{\bf
	\hat{u}_j}) = [\epsilon_{GO}
(\bf{\hat{u}}_i,\bf{\hat{u}}_j)]^{\,\nu} [\epsilon'
(\bf{\hat{r}}_{ij},\bf{\hat{u}}_i,\bf{\hat{u}}_j)]^{\mu}
\end{equation}
\begin{equation}\label{GO}
\epsilon_{GO} ({\bf \hat{u}_i},{\bf \hat{u}_j})=
[1-\chi^2({\bf\hat{u}_i} \cdot {\bf \hat{u}_j})^2]^{-1/2}
\end{equation}

\begin{align}\label{eps}
\epsilon' (\bf{\hat{r}}_{ij},\bf{\hat{u}}_i,\bf{\hat{u}}_j)=
\mbox{~~~~~~~~~~~~~~~~~~~~~~~~~~~~~~~~~~~~~~~~~~~~~~~}
\nonumber \\
\mbox{~~} 1-\frac{\chi'}{2} \left[
\frac{({\bf\hat{r}}_{ij}\cdot{\bf\hat{u}}_i+{\bf\hat{r}}_{ij}\cdot
	{\bf\hat{u}}_j)^2}{1+\chi^{'}{\bf\hat{u}}_i\cdot{\bf\hat{u}}_j}+
\frac{({\bf\hat{r}}_{ij}\cdot{\bf\hat{u}}_i-{\bf\hat{r}}_{ij}\cdot
	{\bf\hat{u}}_j)^2}{1-\chi^{'}{\bf\hat{u}}_i\cdot{\bf\hat{u}}_j}
\right]
\end{align}

\noindent The Kihara factor, $U_K$, is a Lennard-Jones type functional explicitly dependent on the minimum distance between the cores of the interacting particles, $d_m$. This latter distance is determined by the vector joining the centers of mass, with direction $\bf{\hat{r}}_{ij}$ and modulus $r_{ij}$, and the director vectors of the particles, ${\bf \hat{u}}_i$ and ${\bf \hat{u}}_j$  \cite{CUE08}. This results in a body of spherocilindrical shape, given by a cylindrical body of height $L$ and diameter $\sigma$, capped by a hemitoroidal rim with tube radius $L/2$. For this work, we have set $L^*$=$L$/($L$+$\sigma$)=\,0.2. In general, we have used $L+\sigma$ as the unit of length throughout this article. The parameter $\epsilon_0$ scales the strength of the interaction and is formally taken as the unit of energy. In this study, the $U_K$ potential has been truncated at $d_m$=\,3$L$, with an appropriate energy shift to keep it continuous. 

The directional Gay-Berne factor, $\epsilon_{OGB}$, is determined by the three anisotropy constants, $\kappa'$, $\mu$ and $\nu$,  with $\chi^{'}=(\kappa^{'-1/\mu}-1)/(\kappa^{'-1/\mu}+1)$ and $\chi=(L^{*2}-1)/(L^{*2}+1)$. 

A detailed overview of the OGBK model and a rationalization of the choice of parameters that may be employed to modulate the anisotropy of the pair interactions can be found elsewhere \cite{MOR21}. This work considers pair interactions with three qualitatively different orientational dependence. Fig.\,\ref{fig:ogbk} illustrates the interaction models in terms of potential energy curves for reference pair configurations, namely face-face, face-edge, edge-edge parallel and edge-edge crossed. The following cases are specifically considered:

i) Model $K$ ($\kappa'=1$, $\mu=0$ and $\nu=0$): The native discotic Kihara pair potential, dependent only on core distance, without energetic preference for any particular relative pair orientation. 

ii) Model $E$ ($\kappa'=0.1$, $\mu=1$ and $\nu=1$): A pair potential favoring edge-edge parallel interactions,

iii) Model $F$ ($\kappa'=5$, $\mu=1$ and $\nu=1$): A pair potential favoring face-face stacking interactions.

\subsection{\normalsize Simulation Techniques}\label{sec-MC}

Monte Carlo simulations were performed at constant particle number $N$, volume $V$ and temperature $T$ (MC-NVT). These simulations involved systems with $N$=1260 particles and were started from a reference system at high temperature ($T^*$=\,$k_BT/\epsilon_0$=\,10, $k_B$ being the Boltzmann constant), where the formation of stable aggregates was negligible. The systems were then sequentially cooled down, seeking for the thermal onset for particle aggregation, $T^*_c$. At each given temperature, the system was equilibrated with a long simulation run (10$^7$ MC cycles), and averages were subsequently computed over 4$\cdot$10$^6$ MC cycles. One MC cycle consists of $N$ attempts to move (rotation and/or translation) a randomly chosen particle.

The kinetic evolution of the aggregation processes was investigated in Brownian Dynamics (BD) simulations \cite{LOW94,PEI09}. We reproduce here the BD formalism for non-spherical particles succinctly. For a better understanding of the forces and torques of the OGBK model, unpublished to date, their explicit expressions are provided in the Appendix to this work. The position, ${\bf r}_j$, and the orientation, ${\hat{{\bf u}}_{j}}$, of a particle $j$ over time are calculated by the following set of equations \cite{LOW94}:

\begin{align} \label{bd1}
{\bf r}^{\parallel}_j(t+\Delta t) =
{\bf r}^{\parallel}_j(t)+\frac{D_{\parallel}}{k_BT} {\bf F}^{\parallel}_j(t)\Delta t+\\
(2D_{\parallel} \Delta t)^{1/2} R_{\parallel} \textbf{\^{u}$_{j}$}(t) \nonumber \\
\nonumber\\
{\bf r}^{\perp}_j(t+\Delta t) = 
{\bf r}^{\perp}_j(t)+\frac{D_{\perp}}{k_BT} {\bf F}^{\perp}_j(t)\Delta t+ \\
(2D_{\perp} \Delta t)^{1/2} (R^{\perp}_1 \textbf{\^{v}$_{j,1}$}(t)+ R^{\perp}_2 \textbf{\^{v}$_{j,2}$}(t))\nonumber \\
\nonumber\\
\textbf{\^{u}$_{j}$}(t+\Delta t) = 
\textbf{\^{u}$_{j}$}(t)+ \frac{D_{\vartheta}}{k_BT} {\bf T}_j(t)\wedge \textbf{\^{u}$_{j}$}(t)\Delta t+ \\
(2D_{\vartheta} \Delta t)^{1/2} (R^{\vartheta}_1 \textbf{\^{v}$_{j,1}$}(t)+R^{\vartheta}_2 \textbf{\^{v}$_{j,2}$}(t))\nonumber
\end{align}

\noindent where ${\bf r}^{\parallel}_j$ and ${\bf r}^{\perp}_j$ indicate the projections of the position vector of particle  \textit{j} along ${\hat{{\bf u}}_{j}}$ and on the orthogonal plane, respectively; ${\bf F}^{\parallel}_j$ and ${\bf F}^{\perp}_j$ are the parallel and perpendicular components of the forces, respectively; ${\bf T}_j$ is the total torque acting on particle $j$ due to the interactions with neighboring particles of the fluid (see the Appendix for explicit expressions of these forces and torques).  $R^{\parallel}, R^{\perp}_1, R^{\perp}_2, R^{\vartheta}_1$ and $R^{\vartheta}_2$ are independent Gaussian random numbers of variance 1 and zero mean. ${\hat{{\bf v}}_{j,1}}$ and ${\hat{{\bf v}}_{j,2}}$ are two random perpendicular unit vectors, orthogonal to each other and to ${\hat{{\bf u}}_{j}}$. 

\begin{figure*}[t!]
	\centering
	\includegraphics[width=0.85\linewidth]{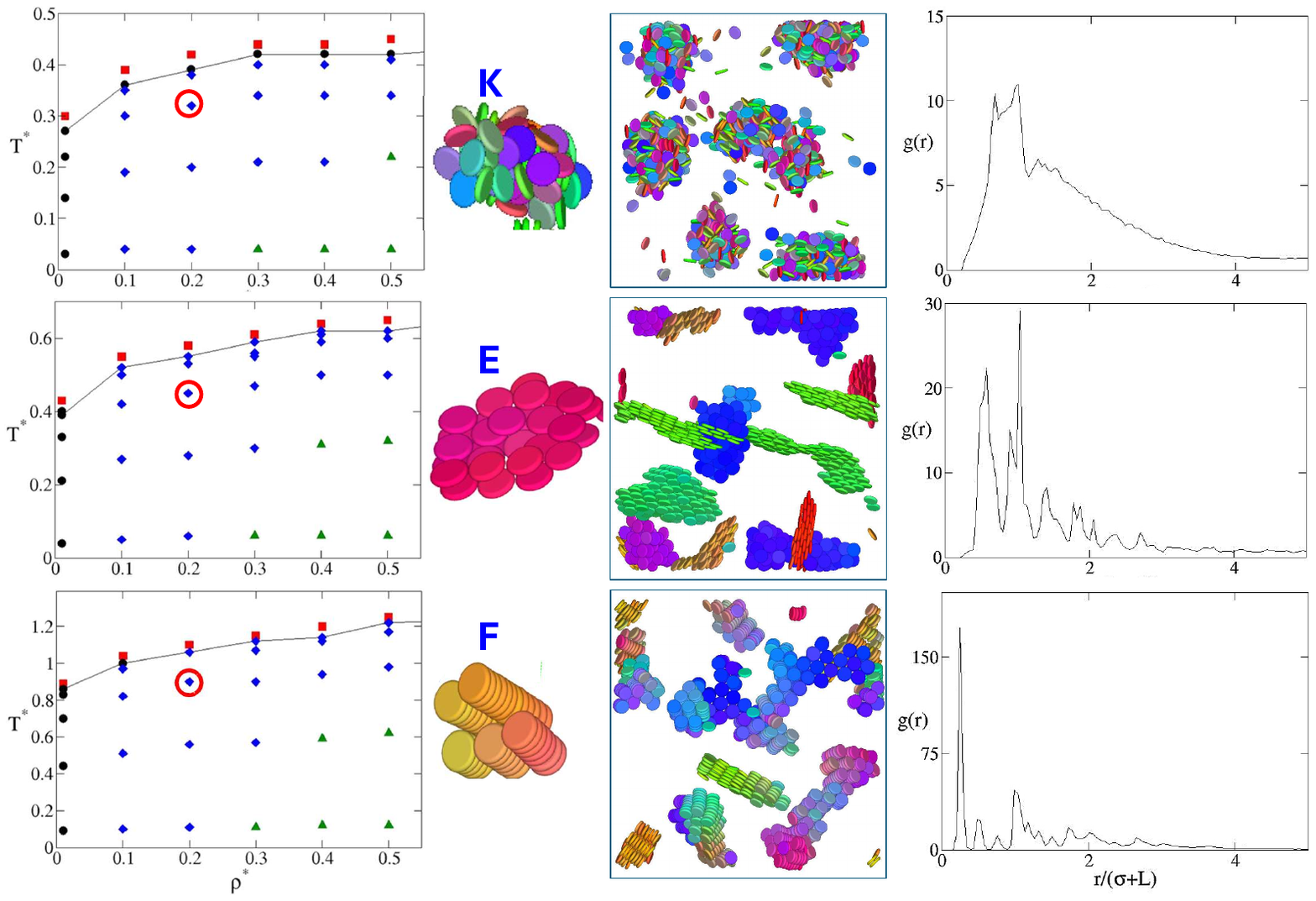}
	\vspace*{-1.0cm}
	\caption{\label{fig:Phase} Classification of the BD simulated states of the K, E and F fluids into aggregation regimes. The solid line represents the onset of aggregation (the highest temperature at which stable clusters are formed). Symbols are used to represent states with no aggregates (\textcolor{red}{$\blacksquare$}), states that coalescence into large clusters (\textcolor{blue}{$ \blacklozenge
			$}),  into multiple small clusters (\textcolor{black}{$\bullet$}), or into a percolated mesoporous structure ({\color{teal}$\blacktriangle$}). Snapshots and radial functions for the states with $\rho^*$=0.2 and T$^*$/$T_c^*$=0.8 are provided to illustrate the internal structure of the clusters and the overall fluid. See Figs.\,\ref{fig-clustersK}--\ref{fig-clustersF} for a more complete set of configurations.} 
\end{figure*}

The short-time diffusion coefficients in the direction parallel and perpendicular to the molecular axis, $D_{\parallel}$, $D_{\perp}$, and the rotational diffusion coefficient, $D_{\vartheta}$, have been calculated with the analytical expressions proposed by Shimizu for spheroids \cite{SHI62}:

\begin{equation}\label{bd3}
D_{\perp}=D_{0}\frac{(2a^{2}-3b^{2})S+2a}{16\pi(a^{2}-b^{2})}b,
\end{equation} 
\begin{equation}\label{bd5}
\tag{14}
D_{\parallel}=D_{0}\frac{(2a^{2}-b^{2})S-2a}{8\pi(a^{2}-b^{2})}b,
\end{equation} 
\begin{equation}\label{bd6}
\tag{15}
D_{\vartheta}=3D_{0}\frac{(2a^{2}-b^{2})S-2a}{16\pi(a^{4}-b^{4})}b.
\end{equation} 
\begin{equation}\label{bd7}
\tag{16}
S=\frac{2}{(b^{2}-a^{2})^{1/2}} \arctan\frac{(b^2-a^2)^{1/2}}{a}
\end{equation}

\noindent where $a$=\,$L$/2, $b$=\,($L$+$\sigma$)/2, $D_0$=\,$\kappa_B T/\mu_s \sigma$ and $\mu_s$ represents the solvent viscosity.

BD simulations were run in the domain of aggregation ($T^*$$<$\,$T^*_c$), reaching up to $t^*$$>$\,6000. Short time steps, $\Delta t^*$$<$\,10$^{-4}$, were applied in order to avoid overlaps without physical meaning. Here, $t^*$=\,$t/\tau$, where $\tau =\sigma^3\mu_s/\epsilon_0$ is the unit of time.
The aim of these lengthy runs was to ensure that the systems reached a stationary state, with a distribution of clusters stable over time at a given density and temperature. It should be remarked that the resulting configurations do not necessarily resemble equilibrium conditions, but may be associated with arrested glassy configurations \cite{POO97,ZAC07,PUE07} or with nucleation of a concentrated phase in the coexistence of two phases \cite{TEN98,AUR05,GHI07}. From the asymptotic configurations so obtained, batches of 100 independent trajectories were run over a span of time of $t^*$=\,100 in order to obtain statistical averages of structural features.

For the purposes of the present work, two particles were considered to be aggregated in the same cluster if their cores were closer than $d_m$$<$\,0.3$L$. After several tests, we  verified that this threshold correctly identified the aggregates for the range of densities here considered. Moreover, a quantitative degree of clustering, $\Theta$, was defined from the number of independent 'entities' in the fluid (monomers or clusters of any size), $N_c$. The average number of particles per entity is then given by P=$N$/$N_c$ ($N$ representing the number of particles in the simulation):

\begin{equation}
\Theta = 1 - \frac{ N_c  }{N} = 1-\frac{1}{ P }
\end{equation}\label{eqr1}

\noindent In the absence of aggregation, the fluid is made of only monomers so that $P$=\,1 and $\Theta$=\,0. Both parameters grow steadily as clusters are formed and increase in size. The limiting values $P$=$N$ and $\Theta$$\sim$\,1 correspond to the collapse of the system into a single cluster or a cross-linked mesoporous arrangement. 

To explore the size distribution of the aggregates, the fraction of particles belonging to an aggregate of size i has been calculated as $F(i) = i \cdot C(i)/N$, where $C(i)$ is the number of clusters of size $i$ in a given configuration.



\section{Results and Discussion}

\begin{figure*}[t]
	\vspace{-0.0cm}
	\begin{center}
		\includegraphics[width=1.5\columnwidth]{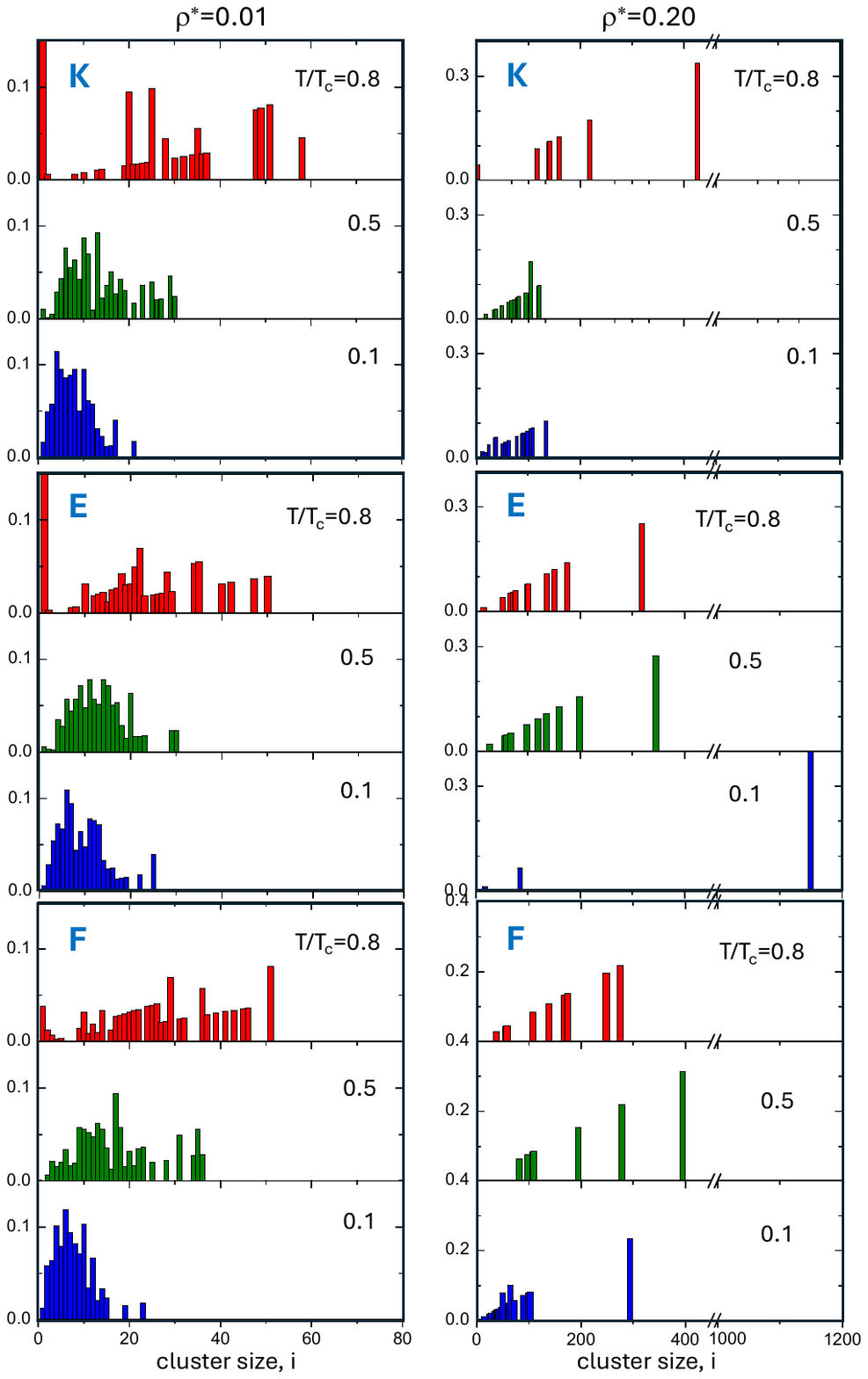}
		\vspace*{-0.5cm} \caption{ Fraction of particles belonging to an aggregate of size $i$, $F(i)$, as defined in Section \ref{sec-MC}. Results for models K, E, and F are shown in the top, middle, and bottom rows, respectively. For each model, the results are shown for densities $\rho^* =0.01$ and $0.2$ (left and right columns, respectively) and temperatures $T/T_c = 0.8, 0.5$, and $0.1$ (red, green and blue bars, respectively). Note the different range of sizes (x-axis) employed in the representations of the two densities.\label{fig-clustersPR}}
	\end{center}
\end{figure*}

\begin{figure*}
	\hspace*{-1.0cm}
	\begin{center}
		\includegraphics[width=15.0cm, angle= 0]{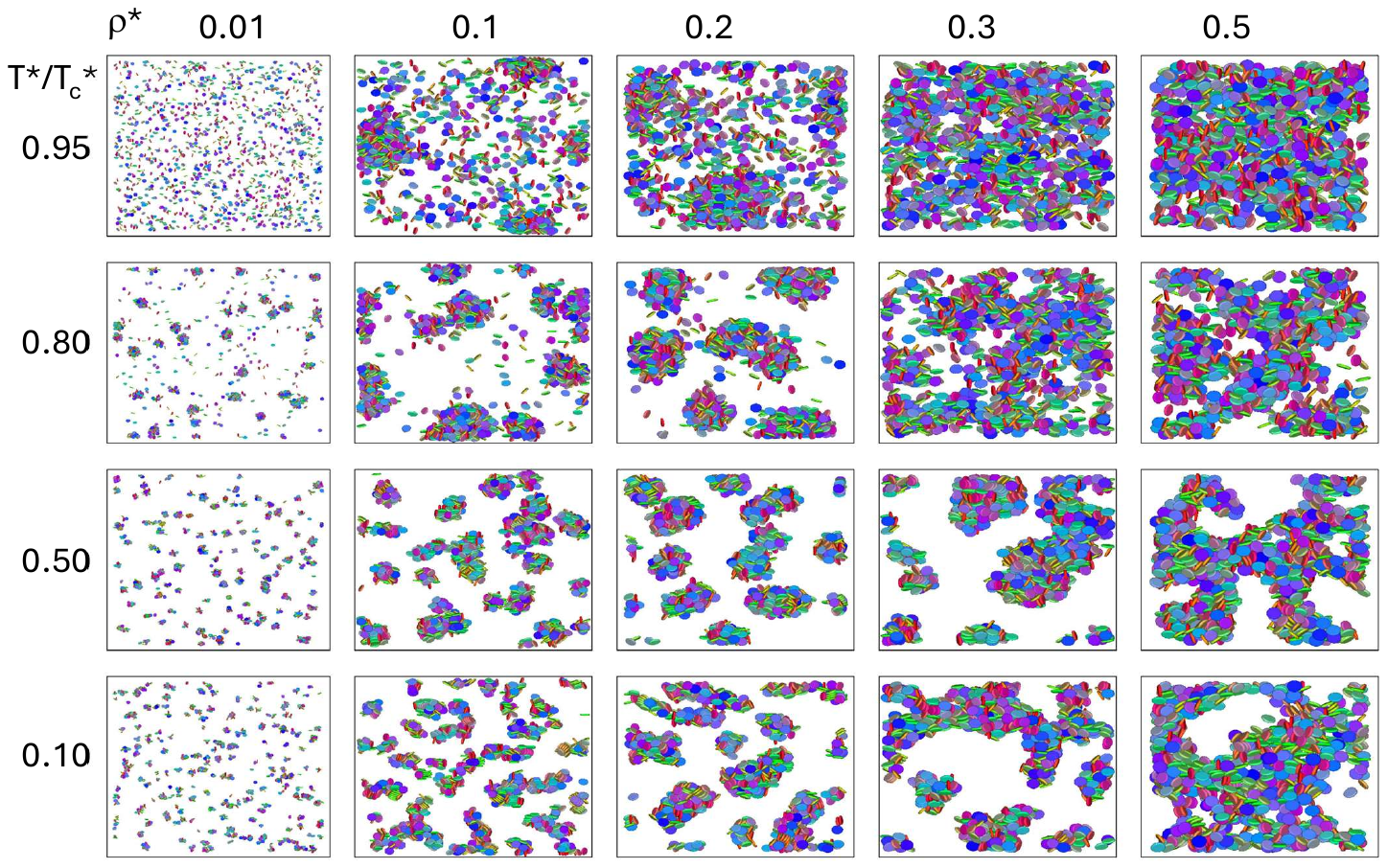}
		\vspace*{-2.7cm} \caption{Examples of snapshots of stable configurations after long simulations, obtained for model K in relevant states. Columns from left to right $\rho^* = 0.01, 0.1, 0.2, 0.3$ and $0.5$. Rows from top to bottom $T^*/T_c^* = 0.95, 0.80, 0.50$ and $0.10$. The color of the particles indicates their orientation.\label{fig-clustersK}}
	\end{center}
\end{figure*}
\begin{figure*}
	\hspace*{-1.0cm}
	\begin{center}
		\includegraphics[width=15.0cm, angle= 0]{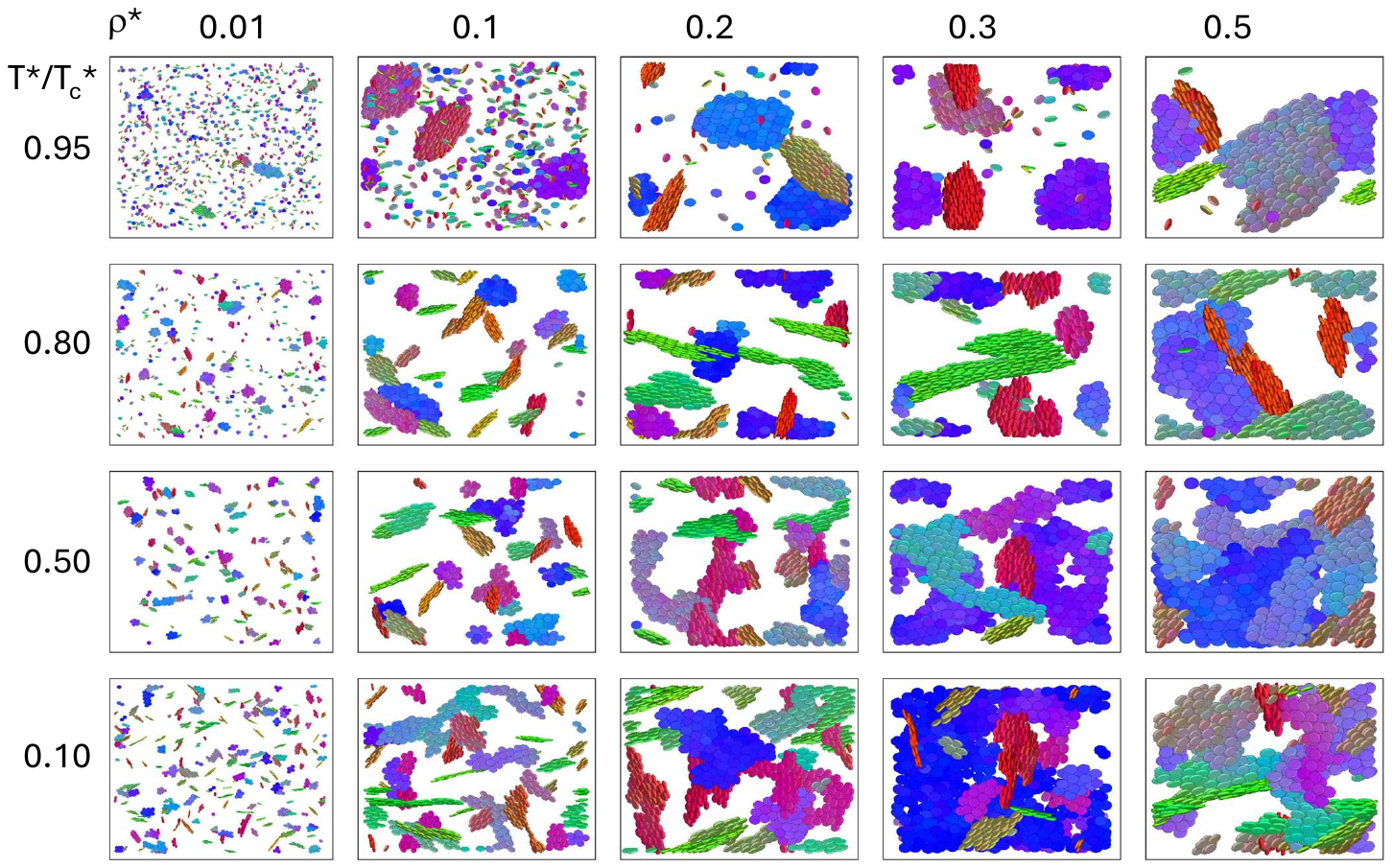}
		\vspace*{-2.7cm} \caption{Examples of snapshots of stable configurations after long simulations, obtained for model E in relevant states. Columns from left to right $\rho^* = 0.01, 0.1, 0.2, 0.3$ and $0.5$. Rows from top to bottom $T^*/T_c^* = 0.95, 0.80, 0.50$ and $0.10$. The color of the particles indicates their orientation.\label{fig-clustersE}}
	\end{center}
\end{figure*}

\begin{figure*}
	\hspace*{-1.0cm}
	\begin{center}
		\includegraphics[width=15.0cm, angle= 0]{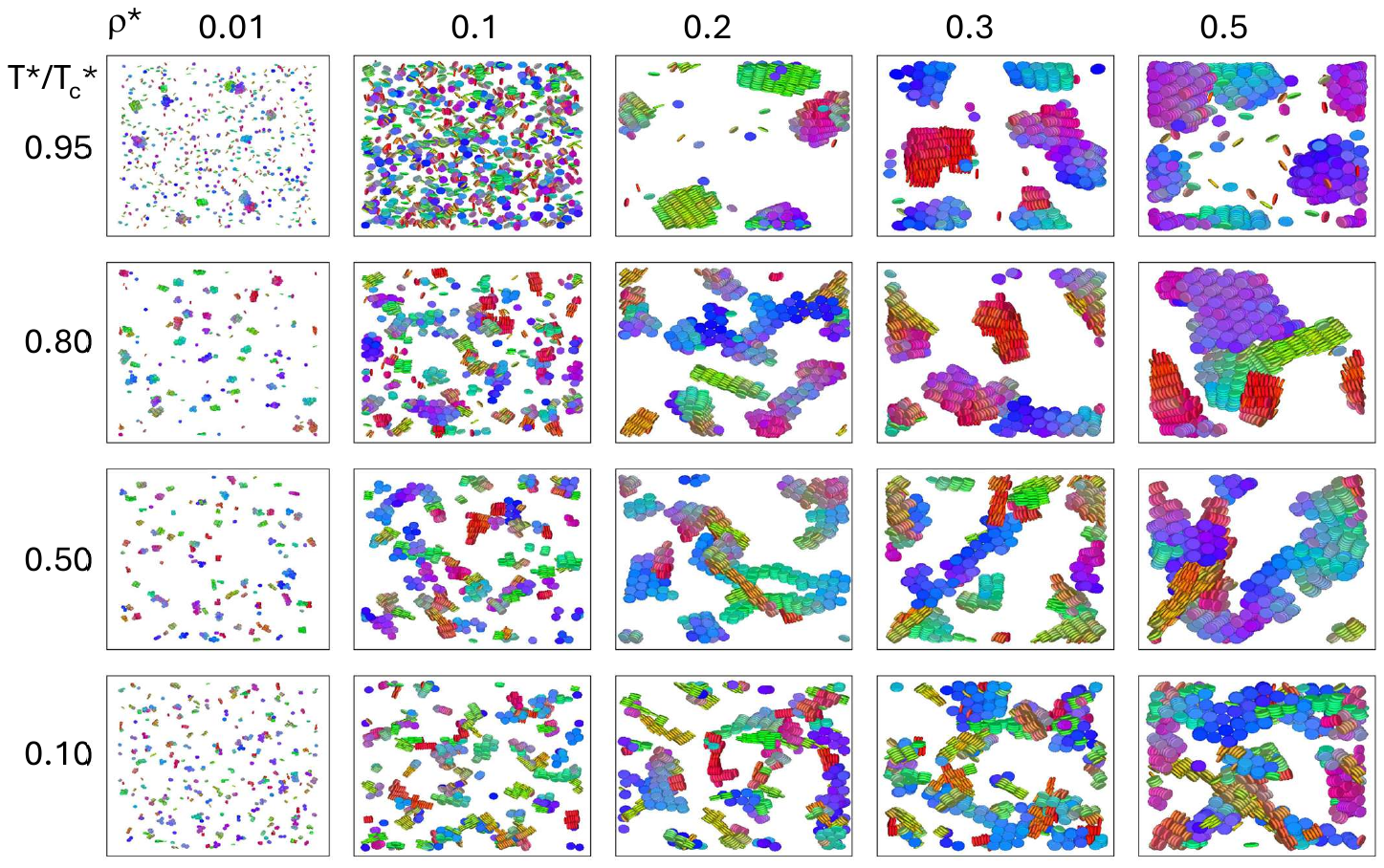}
		\vspace*{-2.7cm} \caption{Examples of snapshots of stable configurations after long simulations, obtained for model F in relevant states. Columns from left to right $\rho^* = 0.01, 0.1, 0.2, 0.3$ and $0.5$. Rows from top to bottom $T^*/T_c^* = 0.95, 0.80, 0.50$ and $0.10$. The color of the particles indicates their orientation.\label{fig-clustersF}}
	\end{center}
\end{figure*}

\subsection{The Aggregation Domain}

Figure \ref{fig:Phase} summarizes in a compact form different aggregation domains observed in the fluid models K, E, and F. Results for each model system are shown for representative states with reduced densities $\rho^*= (N/V)\cdot(L+\sigma)^3 = 0.01-0.5$ and temperatures below the aggregation threshold,  $T/T_c$=\,1.00-0.10. The temperature onsets for particle aggregation, $T_c^*$, obtained from the MC simulations are visualized in Fig.\,\ref{fig:Phase} with a solid line that link the states of highest temperature at which stable aggregates are observed upon isochoric cooling. At temperatures above $T_c^*$, transient clusters are observed mainly as dimers resulting from thermal collisions. At $T_c^*$ and below, a stable distribution of clusters forms beyond kinematic fluctuations.  

It is consistently found that $T_c^*$ increases monotonously with density. For instance, at $\rho^*$=\,0.5, aggregation already initiates at a temperature roughly 1.5 times higher than the one observed at $\rho^*$=\,0.01, which may be considered the high dilution limit. This plausibly follows from the corresponding balance of interaction energy $vs.$ configurational entropy (disaggregated configurations benefit from large system volumes). 

\begin{figure*}[t]
	\centering
	\hspace*{-0.0cm}
	\includegraphics[width=0.7\linewidth]{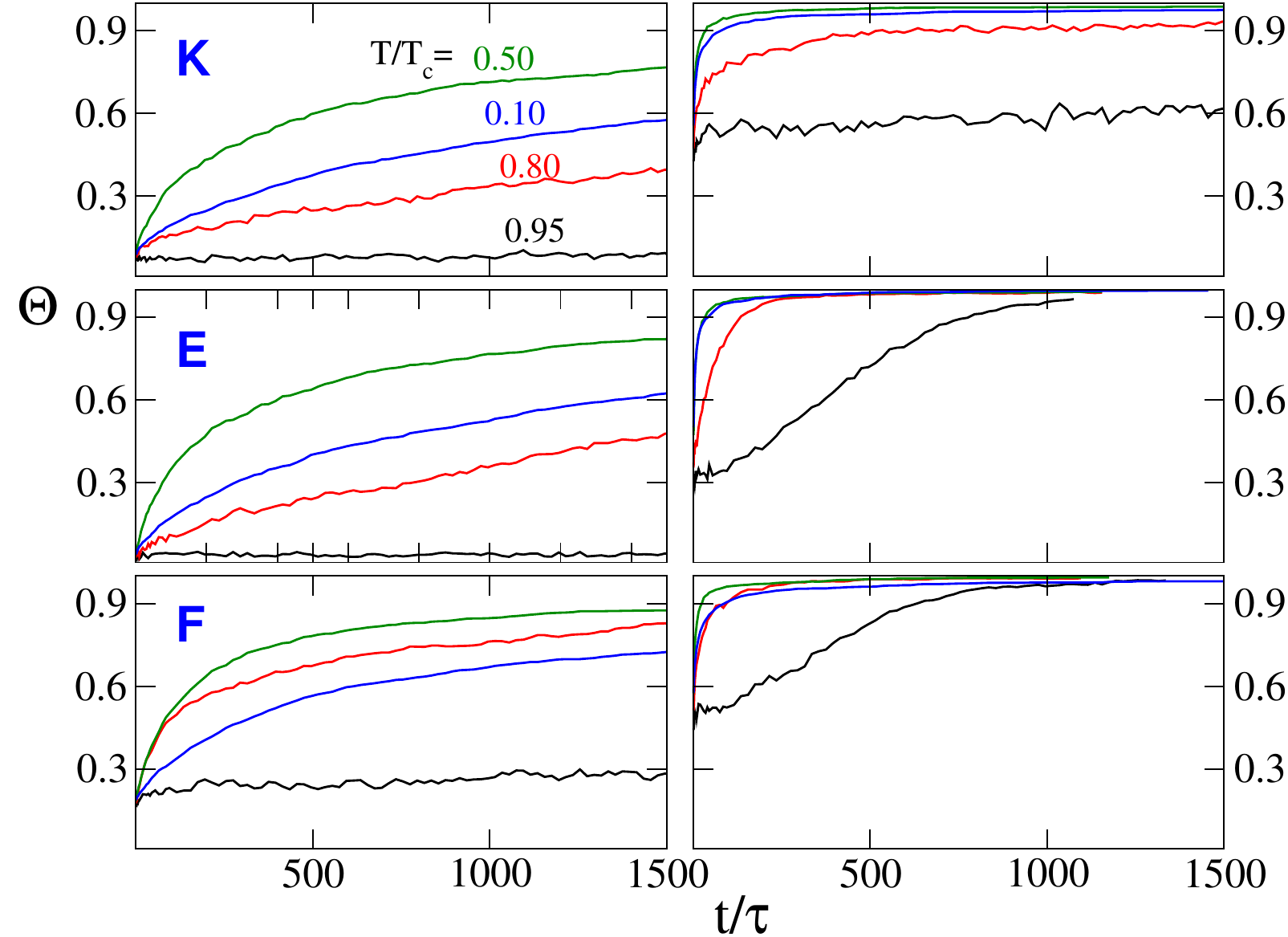}
	\vspace*{-0.0cm}
	\caption{\label{figale1}  Evolution of the degree of clustering, $\Theta$, during the early stages of aggregation. The data are obtained at densities $\rho^*=$0.01 (left column) and 0.2 (right column)  for potentials K,E and F (from top to bottom), at the four indicated temperatures relative to the onset of aggregation $T^*/T^*_c$=\,0.95, 0.80, 0.50 and 0.10 (black, red, green and blue traces, respectively)}    
\end{figure*}

\subsection{Internal structure of the cluster fluids}

A global overview of the internal organization of the fluid states explored in the present study is provided by the cluster size distributions represented in Fig.\,\ref{fig-clustersPR} and the configurational snapshots in Fig.\,\ref{fig-clustersK} (interaction model K), Fig.\,\ref{fig-clustersE} (model E) and Fig.\,\,\ref{fig-clustersF} (model F), which serve to illustrate the variety of cluster organizations found, depending on temperature, density and interaction model. The snapshots shown in these figures come from configurations obtained after long BD simulations. 

For all three models, at the lowest density explored ($\rho^*=0.01$), below $T^*_c$, the clustering phenomenon involves the formation of a large number of small clusters. At higher temperatures, monomers and aggregates coexist, while as $T$ decreases, monomers tend to disappear. For larger densities, $\rho^*\geq$\,0.1, at the higher temperatures, close to $T^*_c$, the three fluids display a propensity to form clusters of large size in coexistence with a bath of particle monomers. At lower temperatures, $T^*$/$T^*_c$$<$\,0.8, aggregation tends to favor the organization of the fluid in an ensemble of polydispersed clusters of smaller size. We argue below that the reduction in cluster size at the lower temperatures correlates with an acceleration of the dynamics of aggregation which then takes place more abruptly. 
At sufficiently high density and low temperature, the individual clusters eventually percolate into a porous mesoscopic structure. 

In the diagrams shown in Fig.\,\ref{fig:Phase}, states are classified according to the size of the clusters. To this end, we have followed the fraction of particles belonging to a cluster of a given size $i$, $F(i)$, defined in Sect. \ref{sec-MC}. This function is represented in Fig.\,\ref{fig-clustersPR}, for some of the states shown in Figs.\,\ref{fig-clustersK} to \ref{fig-clustersF}. The representation of $F(i)$ corroborates that, at low density ($\rho^*=0.01$), a decrease in temperature results in a reduction in the size of the clusters in the fluid. At the higher temperature ($T/T_c=0.8$) the clusters are larger in size but a significant number of monomers still survive. At the lower temperatures ($T/T_c =0.2$ and $0.1$) the fluid is formed by clusters of comparably smaller size, with a negligible presence of free monomers. Fig.\,\ref{fig-clustersPR} also shows that at higher density ($\rho^*=0.2$) the size of the cluster increase and the dependence with temperature becomes more subtle. The clusters reach sizes of a few hundred particles and tend to percolate at low temperature into a single interconnected cluster. While percolation is marked at  $\rho^*=0.2$ only for the E model, it is present for the three models at $\rho^*=0.5$ (see Figs \ref{fig-clustersK}--\ref{fig-clustersF}).

While the above trends are common to the three fluid models, the directionality of the pair interactions has a determining influence on the organization adopted by the particles in the individual aggregates. The internal organization of the individual clusters stabilized in each model fluid is illustrated in Fig.\,\ref{fig:Phase}. The discotic particles interacting with the K potential produce clusters with a roughly globular shape, in which the orientations of the individual particles are weakly correlated mainly through steric interactions. In contrast, potentials E and F lead to clusters with a marked collective positional and orientational order. The edge-edge interactions in the E model consistently give rise to flat aggregates typically organized in multilayer arrangements. The face-face interaction in the F model promotes stacking into columns forming bundles with local hexatic packing. It may seem surprising to find such organization in bundles of columns rather than in longer and more isolated 'wire-like' clusters. The explanation for this behavior can be found in a detailed rationalization of the energetics of this model fluid. Figure 1-d shows that, in model F, the face-face and edge-edge configurations have attractive energy wells of about $-12\epsilon_0$ and $-2\epsilon_0$, respectively. The formation of columnar bundles then results from the multiple pair interactions between particles in neighboring columns. While stacking involves pair interactions with two neighbors, in the bundles a platelet may display edge-edge interactions with up to 12 platelets (if columns are interdigitated). In this way, once the columns become sufficiently long, the formation of columnar bundles competes with the growing of the individual columns. 

The radial distribution functions for the three types of cluster geometry are shown in Fig.\,\ref{fig:Phase}. The characteristic strong short-range correlations are evident in the radial functions of models E and F. For model E, the radial distribution shows a peak at a distance $r^*=1$, indicating edge-to-edge contact. Peaks at shorter distances are indicative of multilayered arrangements, as shown in Fig.\,\ref{fig:Phase}. In contrast, the face-to-face configuration of model F favors the peak in the radial distribution at a distance $r^*$=\,$L^*$=\,0.2,  while that of the K model displays a broader structure, though still with significant correlations at $r^*$$<$\,1 . The structural features of the clusters just described emerge at initial stages of aggregation and then drive the subsequent growth of the clusters and their eventual interconnection, leading to porous structures with differentiated internal topologies.

\subsection{Aggregation kinetics}

From a kinetic perspective, we identified a non-monotonous change with temperature of the rate of aggregation during the early stages of cluster formation. In general terms, clustering proceeds at a slow pace at temperatures close to $T_c^*$. The rate of cluster formation increases upon cooling, although this trend reverses at sufficiently low temperatures.
Fig.\,\ref{figale1} illustrates this behavior through the evolution of the degree of clustering $\Theta$  obtained in the BD simulations for densities $\rho^*$=\,0.01 and 0.2. We recall that our simulations start from a configuration of monomers ($\Theta$=\,0) and aggregation is induced by imposing a temperature below $T_c^*$. The kinetics of clustering is then driven by the pair interaction energy and the rate of diffusion of the particles potentially incorporating to clusters. 

At temperatures immediately below the aggregation threshold, $T/T_c=0.95$, clusters form at slow rate in the three fluid models. Even if the asymptotic degree of aggregation approaches $\Theta$$\sim$\,1, it takes a substantial time for clusters to build up, $e.g.$ still more than $t/\tau$=\,1000 for the $\rho^*=\,0.2$ states depicted in Fig.\,\ref{figale1}. At lower temperatures, the systems enter a regime of faster aggregation rates. The largest pace of aggregation is observed in our simulations at $T/T_c=0.50$, while at lower temperatures it eventually slows down due to the smaller diffusion rates of the particles, as exemplified in Fig.\,\ref{figale1} for $T/T_c=0.10$.

The combined analysis of the structural and kinetic features of the clustering processes indicates that the slower paces of aggregation correlate with a trend for the particles to concentrate around a reduced number of nucleation centers, yielding clusters of larger size. As the temperature is lowered, the aggregation rates are overall enhanced, which results in a more abrupt collapse of the particles into multiple nucleation centers, leading to clusters of comparatively smaller size.

\section{Final remarks}

The aim of the study has been to provide a modeling framework to investigate thermotropic aggregation processes in fluids of discotic particles. The OGBK coarse-grained model has served to explore the effect of the directionality of the pair interactions on the topology of clustered structures produced upon the sudden cooling of an initially disaggregated fluid.

The phenomenology exposed by the present simulation can be expected to resemble qualitatively the behavior found in colloidal and macromolecular systems based on planar molecules. For instance, the occurrence of the slow and fast aggregation regimes described here as a function of the clustering temperature can be related to the similar regimes found for different colloids with attractive pair interactions \cite{WEI85,LIN89,DEH01}. Moreover, the percolated arrangements stabilized at the higher densities and lower temperatures are in concordance with the ones found in experiments of attractive glasses and gels \cite{ZAC07,RUZ11}. In general terms, the anisotropic structural features of the clusters resulting from specific interactions should aid in the rationalization of the formation of flocculates and crystallites in contexts of synthetic and natural (bio)macromolecular sciences \cite{MUL08,NEH18,WIL21}).

An important aspect that is admittedly not fully addressed in this work is whether the observed aggregation phenomena are the consequence of scenarios of stable cluster fluids or are linked to vapor-liquid or vapor-solid coexistence regimes \cite{LU06}. In fact, a recent publication shows that the aggregation region exposed in for model K (the native Kihara fluid) falls below the liquid-vapor coexistence curve \cite{CUE24}. The present simulations would then be probing the kinetics of a transition from a high-temperature vapor to a low-temperature gas-liquid coexistence. The situation is uncertain for the E and F fluids, since liquid-vapor coexistence has not been described for these models. Hence, questions related to the metastability of the clusters currently described remain open and pending for future studies, ideally incorporating direct-coexistence and free-energy approaches.

\begin{acknowledgments}
We acknowledge funding from the Government of Spain (Project TED2021-130683B-C21). We are indebted to the High-Performance Computing Center C3UPO at Universidad Pablo de Olavide for technical assistance. 
\end{acknowledgments}

\section*{Data Availability Statement}

The data that support the findings of this study are available from the corresponding authors upon reasonable request.

\appendix

\section{Appendix I. Forces and torques for the OGBK interaction model}

We provide here explicit expressions for the forces and torques of the OGBK interaction model (equations \ref{OGBK} to \ref{eps}), which remain unpublished to date. The framework is similar to the one develop for the Kihara model in one of Dr. Carlos Vega's early works. \cite{VEG90}.

The force ${\bf F}_{ij}$ associated with two particles with OGBK interactions can be developed into the following set of expressions (see Eq.\,\ref{OGBK}):

\begin{align}\label{ape1}
{\bf F}_{ij} = -\vec{\nabla}_{{\bf r}_{ij}} U_{OGBK}({\bf r_{ij}},{\bf \hat{u}_i},{\bf\hat{u}_j})= \nonumber\\\
= -\epsilon_{OGB} ({\bf \hat{r}_{ij}},{\bf \hat{u}_i},{\bf\hat{u}_j}) \vec{\nabla}_{{\bf r}_{ij}}U_{K}(d_m)\nonumber\\ 
- U_{K}(d_m) \vec{\nabla}_{{\bf r}_{ij}} \epsilon_{OGB} ({\bf \hat{r}_{ij}},{\bf \hat{u}_i},{\bf\hat{u}_j})\nonumber\\
\end{align}

\noindent where the operator $\vec{\nabla}_{{\bf r}_{ij}}$ represents the gradient with respect to the relative position, ${\bf r}_{ij}$=\,${\bf r}_{j}$-${\bf r}_{i}$. 


The first term in Eq.\,\ref{ape1} includes the contribution of the Kihara part of the potential to the force. It can be calculated according to the expression developed by Vega and Lago \cite{VEG90}:

\begin{align}\label{ape3}
\vec{\nabla}_{{\bf r}_{ij}}U_{K}(d_m) =\frac{d(U_{K})}{d(d_m)}·\frac{{\bf d}_{m}}{d_m}\nonumber\\
=\epsilon_0\left[ \frac{24}{d_m^8}-\frac{48}{d_m^{14}}\right]{\bf d}_{m}\nonumber\\
\end{align}

\noindent with ${\bf{d}}_{m}$ a vector in the direction of the shorter distance from {i} to {j}\,\cite{CUE08}.

The second term in Eq.\,\ref{ape1} is given by the following expressions, which take into account that $\vec{\nabla}_{{\bf r}_{ij}}\gamma=0$:

\begin{align}\label{ape5}
\vec{\nabla}_{{\bf r}_{ij}} \epsilon_{OGB} ({\bf \hat{r}_{ij}},{\bf \hat{u}_i},{\bf\hat{u}_j})= \frac{\partial{\epsilon_{OGB}}}{\partial\alpha}\vec{\nabla}_{{\bf r}_{ij}}\alpha+\frac{\partial{\epsilon_{OGB}}}{\partial\beta}\vec{\nabla}_{{\bf r}_{ij}}\beta
\end{align}
\begin{align}\label{ape6}
\frac{\partial{\epsilon_{OGB}}}{\partial\alpha}&=&-\chi^{'}\mu\frac{\epsilon_{OGB}}{\epsilon'}\frac{\alpha}{1+\chi^{'}\gamma}\nonumber\\
\frac{\partial{\epsilon_{OGB}}}{\partial\beta}&=&-\chi^{'}\mu\frac{\epsilon_{OGB}}{\epsilon'}\frac{\alpha}{1-\chi^{'}\gamma}\nonumber\\
\vec{\nabla}_{{\bf r}_{ij}}\alpha&=&\frac{{\bf \hat{s}}-\alpha{\bf \hat{r}_{ij}}}{r_{ij}}\nonumber\\
\vec{\nabla}_{{\bf r}_{ij}}\beta&=&\frac{{\bf \hat{s'}}-\beta{\bf \hat{r}_{ij}}}{r_{ij}}\nonumber\\
\end{align}
\begin{align}\label{ape4}
\alpha &=& {\bf \hat{r}_{ij}}\cdot({\bf \hat{u}_i}+{\bf\hat{u}_j}) = {\bf \hat{r}_{ij}}\cdot{\bf \hat{s}}\nonumber\\
\beta &=& {\bf \hat{r}_{ij}}\cdot({\bf \hat{u}_i}-{\bf\hat{u}_j}) = {\bf \hat{r}_{ij}}\cdot{\bf \hat{s'}}\nonumber\\
\gamma&=&{\bf \hat{u}_i}\cdot{\bf\hat{u}_j}\nonumber\\
\end{align}

Analogously, the torques associated with the OGBK forces can be outlined as:

\begin{equation}\label{ape7}
{\bf T}_{ij} =\epsilon_{OGB} ({\bf \hat{r}_{ij}},{\bf \hat{u}_i},{\bf\hat{u}_j}){\bf T}^{K}_{ij}
+U_{K}(d_m){\bf T}^{OGB}_{ij}
\end{equation}

\noindent being ${\bf T}_{ij}$, ${\bf T}^{K}_{ij}$ and ${\bf T}^{OGB}_{ij}$ the torque on particle $j$ due the interaction with particle $j$, and the contribution from the Kihara and GB terms of the potential, $U_K$ and $\epsilon_{OGB}$, respectively. For the calculation of ${\bf T}^{K}_{ij}$ we start with the formal definition of the torque on a particle $j$ due to the force exerted by the particle $i$ \cite{ALL89,FIN93}:

\begin{equation}\label{ape8}
{\bf T}^{K}_{ij} = {\bf r}^{\alpha}_j\wedge {\bf F}^K_{ij} = - {\bf r}^{\alpha}_j\wedge \vec{\nabla}_{{\bf r}_{ij}}U_{K}(d_m)
\end{equation}

\noindent Here, ${\bf r}^{\alpha}_j$ denotes the vector from the center of particle $j$ to the point of application of the force ${\bf F}^K_{ij}$, which coincides with the closest point in $j$ to particle $i$ and can be calculated using the algorithm outlined in ref.\,\cite{CUE08}. A fundamental difference with respect to the fluids of rod-like particles studied in refs.\,\cite{ALL89,FIN93} is that, in disk-shaped particles, ${\bf r}^{\alpha}_j$ is perpendicular to the direction vector ${\bf\hat{u}_j}$. Therefore, when substituting in the dynamic equation (Eq.\,\ref{bd1}), we arrive at 

\begin{align}\label{ape9}
{\bf T}^{K}_{ij} \wedge {\bf\hat{u}_j}&=& {\bf F}^K_{ij}({\bf r}^{\alpha}_j\cdot{\bf\hat{u}_j}) - {\bf r}^{\alpha}_j ({\bf\hat{u}}_j\cdot{\bf F}^K_{ij})=\nonumber\\
&=&- {\bf r}^{\alpha}_j ({\bf\hat{u}}_j\cdot{\bf F}^K_{ij})\nonumber\\
&=& + \epsilon_0\left[ \frac{24}{d_m^8}-\frac{48}{d_m^{14}}\right]({\bf d}_{m}\cdot{\bf\hat{u}}_j){\bf r}^{\alpha}_j\nonumber\\
\end{align}

The contribution to the torque of the Gay-Berne contribution to the torque can be calculated as \cite{ALL06}

\begin{align}\label{ape10}
{\bf T}^{OGB}_{ij} = - {\bf\hat{u}}_j \wedge \vec{\nabla}_{{\bf\hat{u}}_j} \epsilon_{OGB}
\end{align}
\begin{align}\label{ap11}
\vec{\nabla}_{{\bf\hat{u}}_j} \epsilon_{OGB} ({\bf \hat{r}_{ij}},{\bf \hat{u}_i},{\bf\hat{u}_j})=\nonumber\\ \frac{\partial{\epsilon_{OGB}}}{\partial\alpha}\vec{\nabla}_{{\bf\hat{u}}_j}\alpha+\frac{\partial{\epsilon_{OGB}}}{\partial\beta}\vec{\nabla}_{{\bf\hat{u}}_j}\beta
+\frac{\partial{\epsilon_{OGB}}}{\partial\gamma}\vec{\nabla}_{{\bf\hat{u}}_j}\gamma\nonumber\\
\end{align}

For the OGBK model (equations \ref{OGBK} to \ref{eps}), the following explicit expressions are obtained:

\begin{align}\label{ape12}
\frac{\partial{\epsilon_{OGB}}}{\partial\gamma}=\nu\frac{\epsilon_{OGB}}{\epsilon_{GO}}\frac{\partial\epsilon_{GO}}{\partial\gamma}+\mu\frac{\epsilon_{OGB}}{\epsilon'}\frac{\partial\epsilon'}{\partial\gamma}\nonumber\\
\frac{\partial\epsilon_{GO}}{\partial\gamma} = \chi^2\gamma\epsilon^3_{GO}\nonumber\\
\frac{\partial\epsilon'}{\partial\gamma}=\frac{\chi^{'2}}{2}\left(\frac{\alpha^2}{(1+\chi^{'}\gamma)^2}- \frac{\beta^2}{(1+\chi^{'}\gamma)^2}   \right)\nonumber\\
\vec{\nabla}_{{\bf\hat{u}}_j}\alpha={\bf \hat{r}_{ij}}\nonumber\\
\vec{\nabla}_{{\bf\hat{u}}_j}\beta=-{\bf \hat{r}_{ij}}\nonumber\\
\vec{\nabla}_{{\bf\hat{u}}_j}\gamma={\bf \hat{e}_{i}}\nonumber\\
\end{align}

Finally, the rotational BD equation (Eq.\,\ref{bd1}) can be solved with:

\begin{align}\label{ape13}
{\bf T}^{OGB}_{ij}\wedge {\bf\hat{u}}_j &=& - ({\bf\hat{u}}_j \wedge \vec{\nabla}_{{\bf\hat{u}}_j} \epsilon_{OGB})\wedge {{\bf\hat{u}}_j}\nonumber\\
&=&-\vec{\nabla}_{{\bf\hat{u}}_j} \epsilon_{OGB}+{\bf\hat{u}}_j(\vec{\nabla}_{{\bf\hat{u}}_j} \epsilon_{OGB}\cdot{\bf\hat{u}}_j)\nonumber\\
\end{align}

\nocite{*}
\bibliography{ogbkbib.bib}

\end{document}